\documentclass{eas}
\usepackage{graphicx}
\newcommand{\msun}{$\mathrm{M_{\odot}}$}

\begin{document}

\title{Lithium abundance evolution in open clusters: Hyades, NGC752, and M67} 
\author{M. Castro}\address{Departamento de F\'isica Te\'orica e Experimental (DFTE), Universidade Federal do Rio Grande do Norte (UFRN), Natal (RN), Brazil}
\author{G. Pace}\address{Centro de Astrof\'isica da Universidade do Porto (CAUP), Porto, Portugal}
\author{J.-D. do Nascimento Jr.}\sameaddress{1}
\runningtitle{Castro \etal: Lithium abundance evolution in open clusters}
\begin{abstract}
Mixing mechanisms bring the Li from the base of the convective zone to deeper and warmer layers where it is destroyed. These mechanisms are investigated by comparing observations of Li abundances in stellar atmospheres to models of stellar evolution. Observations in open  cluster are especially suitable for this comparison, since their age and metallicity are homogeneous among their members and better determined than in field stars. In this  work, we compare the evolution of Li abundances in three different clusters: the Hyades, NGC752, and M67. Our models are calculated with microscopic diffusion and transport of chemicals by meridional circulation, and calibrated on the Sun. These comparisons allow us to follow the evolution of Li abundance as a function of stellar mass in each cluster and as a function of the age by comparing this evolution in each cluster. We evaluate the efficiency of the mixing mechanisms used in the models, and we try to identifiy the lacking mechanisms to reproduce the observed evolution of Li abundance.   
\end{abstract}
\maketitle
\section{Introduction}
One of the challenging question in the modern stellar physics is to know if the Sun has a peculiar abundance of Li regarding to other stars of the same type. The present day solar Li abundance (A(Li) = 1.05) is much lower than the meteoritic Li abundance (A(Li) = 3.26; Asplund {\em et al.\/} \cite{asplund2009}). Standard models fail to reproduce solar Li destruction during the main sequence evolution.

The most recent observations show solar twins exhibiting lithium abundances comparable to the solar value: HIP 56948: A(Li) $= 1.28 \pm 0.05$ (Mel\'endez {\em et al.\/} \cite{melendez2012}); HIP 73815: A(Li) $< 0.90 \pm 0.20$ (Mel\'endez \& Ram\'irez \cite{melendez&ramirez2007}); YPB637, YPB1194, and YPB1787 in the open cluster M67: A(Li) $= 1.5$, $<1.3$, and $1.6$, respectively (Castro {\em et al.\/} \cite{castro2011}); CoRoT ID102684698: A(Li) $= 0.85 \pm 0.35$ (do Nascimento {\em et al.\/} \cite{donascimento2013}).

Those observations and the improvements of evolution models tend to show that solar Li abundance is not peculiar but a product of depletion due to non-standard mixing which affects both the Sun and the solar twins. Many authors have studied different mixing processes that could cause Li depletion, such as internal gravity waves (Montalban \& Schatzmann \cite{montalban&schatzman2000}; Charbonnel \& Talon \cite{charbonnel&talon2005}), overshooting mixing (Xiong \& Deng \cite{xiong&deng2009}), and meridional circulation (do Nascimento {\em et al.\/} \cite{donascimento2009}). All these studies suggest that Li depletion is a function of several parameters such as mass, age, metallicity, and angular momentum history (Charbonnel \& Talon \cite{charbonnel&talon2005}; do Nascimento {\em et al.\/} \cite{donascimento2009}; Meléndez {\em et al.\/} \cite{melendez2010}; Baumann {\em et al.\/} \cite{baumann2010}; Castro {\em et al.\/} \cite{castro2011}).

We studied the evolution of the Li abundance in three open clusters of different ages (Hyades, NGC752, and M67). We  collected all  published  data  to the  best  of our  knowledge, re-analysed them and compared with TGEC evolution models calibrated on the Sun. In Sec. \ref{sec_models}, we present the physics in our models. In Secs. \ref{sec_M67}, \ref{sec_NGC752}, and \ref{sec_Hyades}, we compare our models to the observations for the open clusters M67, NGC752, and Hyades, respectively. In Sec. \ref{sec_Conclusions}, we announce some conclusions. 

\section{TGEC models}
\label{sec_models}

The Toulouse Geneva Evolution Code (TGEC) compute 1D evolution models from ZAMS to the top of the RGB. Details about the input physics and the mixing processes in the models can be found in Pace {\em et al.\/} \cite{pace2012}.  

With the assumption that the Sun is a common star among solar-type stars, we chose to calibrate our models on the Sun. We used the method described in Richard {\em et al.\/} \cite{richard2004}. We obtained at the solar age, a Li abundance $A(\mathrm{Li}) = 1.04$.

\section{Li in the open cluster M67}
\label{sec_M67}

Our data base is composed of 103 stars from a compilation of literature sources of Li abundance measurements, namely Canto Martins {\em et al.\/} \cite{cantomartins2011}, Castro {\em et al.\/} \cite{castro2011}, Pasquini {\em et al.\/} \cite{pasquini2011}, Randich {\em et al.\/} \cite{randich2007}, Jones {\em et al.\/} \cite{jones99}, and Balachandran \cite{balachandran95}. 

\begin{figure}
  \centering
  \includegraphics[width=5.5cm]{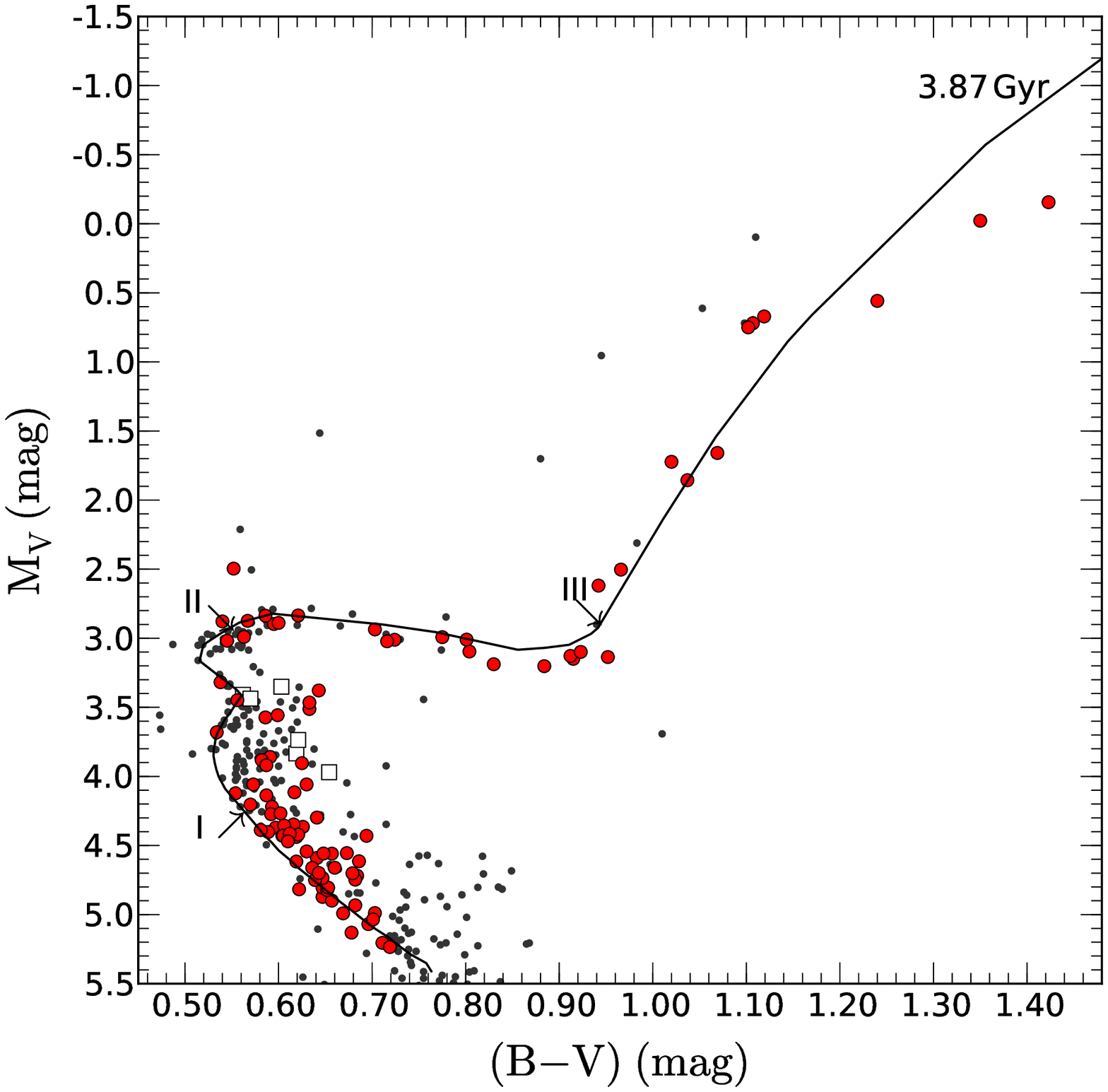}
  \qquad
  \includegraphics[width=5.8cm,height=5.4cm]{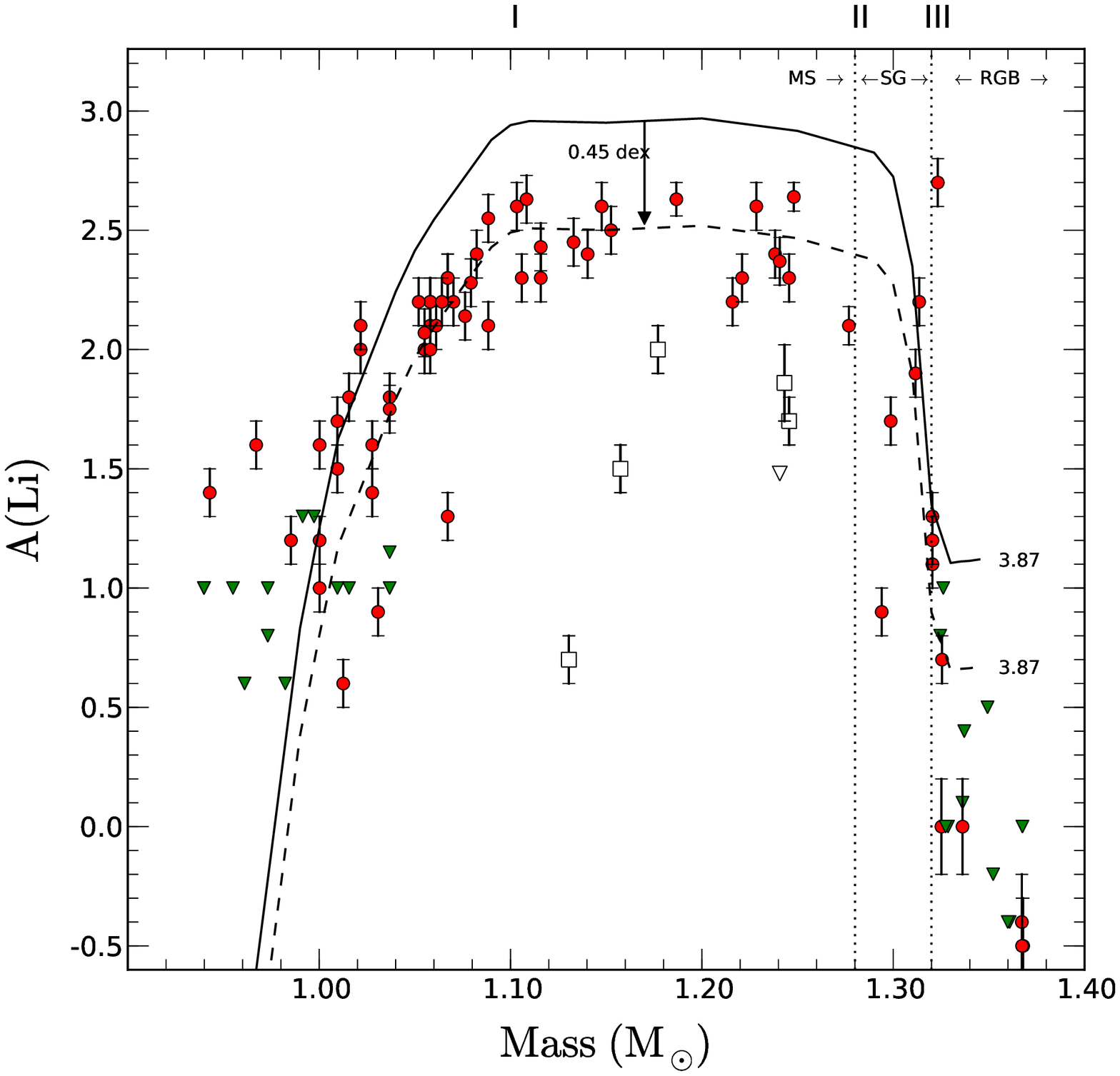}
  \caption{\textit{Left}: Color-magnitude diagram of the open cluster M67. \textit{Right}: Li abundance as a function of stellar mass for the open cluster M67. From Pace \etal \ \cite{pace2012}.}
  \label{fig_M67}
\end{figure}

In the left panel of Figure \ref{fig_M67}, we show the color-magnitude diagram of the open cluster M67 (Pace {\em et al.\/}, 2012), obtained with a reddening $E(B - V) = 0.02$ mag and a distance modulus $(m - M)_0 = 9.68$ mag. The isochrone of age 3.87 Gyr was calculated with models of metallicity [Fe/H] = 0.01 and masses from 0.90 to 1.31 \msun \ with a step of 0.01 \msun.

The right panel of Figure \ref{fig_M67} shows the Li abundance as a function of mass. The masses of the observational points were determined by interpolation in the CMD with the isochrone.

For stars with mass lower than 1.10 \msun (until the point I in Fig. \ref{fig_M67}), the Li depletion progressively increases toward lower masses. A dispersion is observed in this mass range and confirmed by several authors (Pasquini {\em et al.\/} \cite{pasquini97}; Jones {\em et al.\/} \cite{jones99}; Randich {\em et al.\/} \cite{randich2002},\cite{randich2007}; Pasquini {\em et al.\/} \cite{pasquini2008}), probably due to the history of angular momentum (Charbonnel \& Talon \cite{charbonnel&talon2005}).

Stars with masses between 1.10 \msun \ and 1.28 \msun \ (from I to II) are close to the turn off (or about to leave the MS). The surface convection zone and the mixed layers below the convection zone are too thin to reach temperatures high enough to cause the nuclear destruction of Li, whose abundance remains large. It exists an offset of 0.45 dex between our models and observations, showing an insufficient destruction of Li by rotationally driven mixing mechanisms.

Stars with mass larger than 1.28 \msun \ (from II to III) are sub-giants following an evolutionary path from the turn-off to the RGB. This range of masses corresponds to the Li-dip of M67 (Balachandran \cite{balachandran95}).

\section{Li in the open cluster NGC752}
\label{sec_NGC752}

For the open cluster NGC752, we adopted the Li equivalent width measurements from the 2.1 m and 4 m telescopes of Kitt Peak National Observatory (Hobbs \& Pilachowsky \cite{hobbs&pilachowsky86}; Pilachowsky \& Hobbs \cite{pilachowsky&hobbs88}; Pilachowsky \etal \ \cite{pilachowsky88}) and the 3.58 m SARG@TNG (Sestito {\em et al.\/} \cite{sestito2004}). We used the photometry from Daniel {\em et al.\/} (\cite{daniel94}). We found two different metallicities in the literature: [Fe/H] $= -0.15 \pm 0.05$ dex obtained from Daniel {\em et al.\/} (\cite{daniel94}), and [Fe/H] $\sim +0.10$ dex inferred by high-resolution spectroscopy of 4 giants by Carrera \& Pancino (\cite{carrera&pancino2011}).
 
\begin{figure}
  \centering
  \includegraphics[width=5.5cm,height=5.5cm]{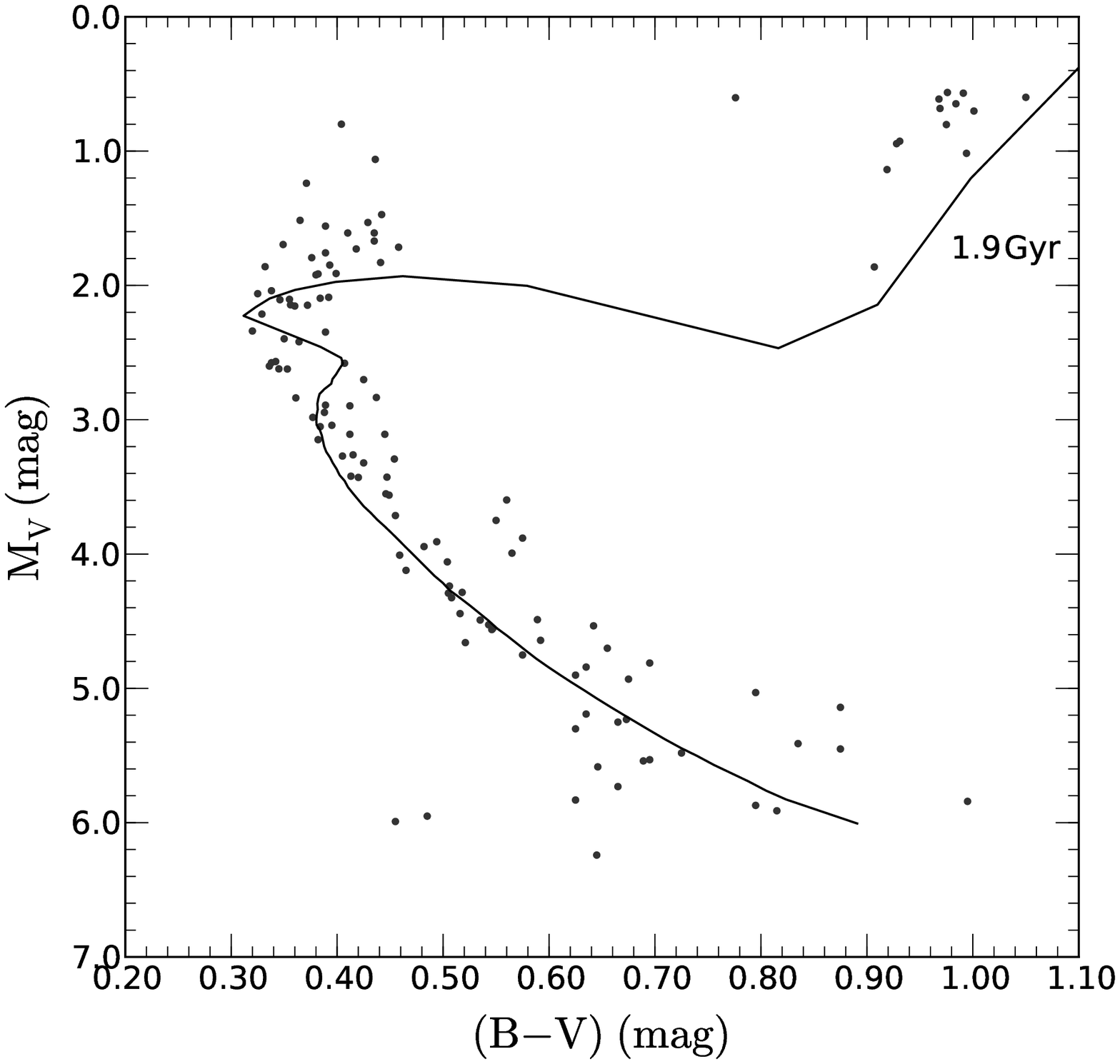}
  \qquad
  \includegraphics[width=5.7cm,height=5.5cm]{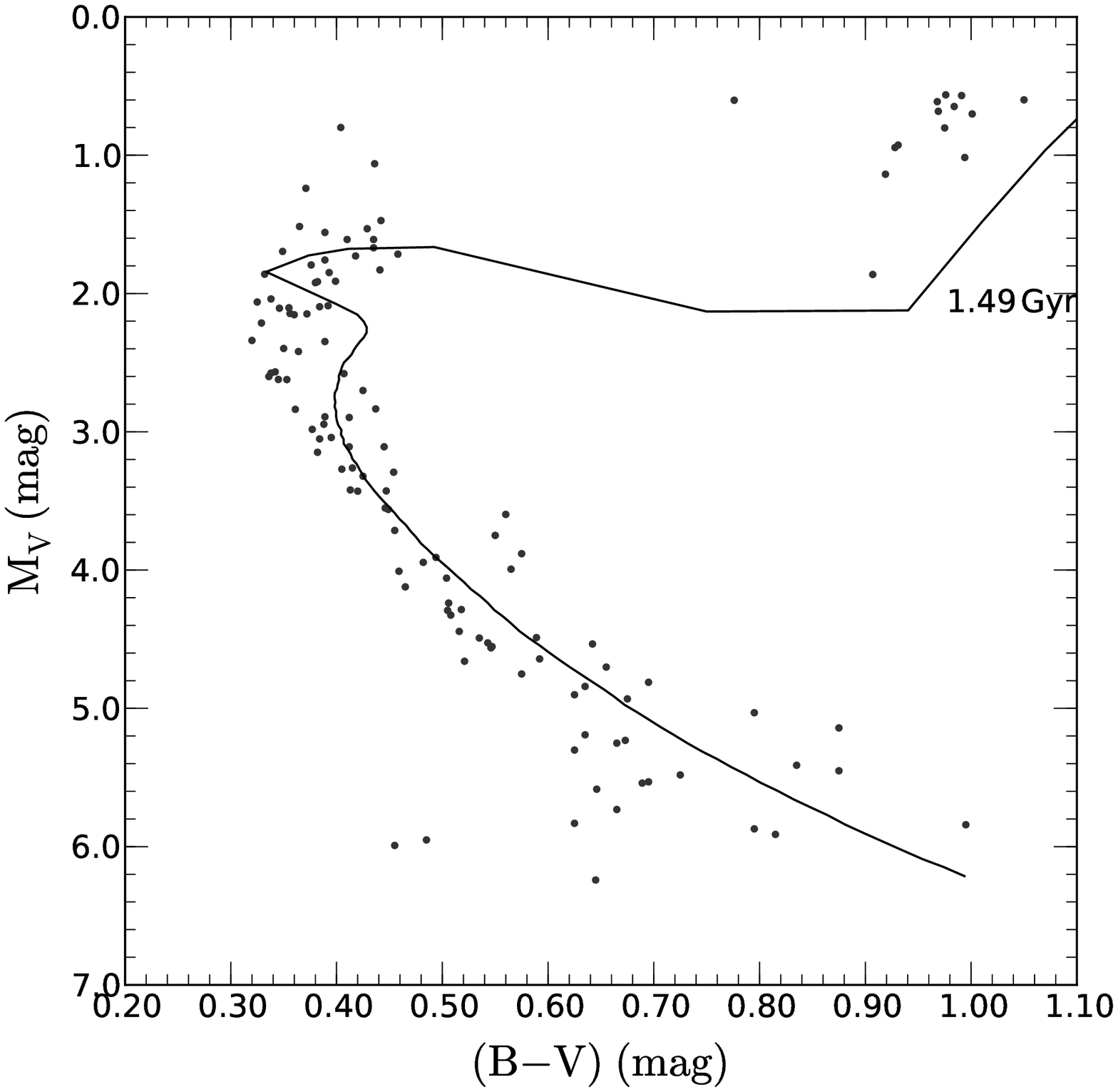}
  \caption{Color-magnitude diagram of the open cluster NGC752 for metallicities [Fe/H] = -0.15 (\textit{left}) and [Fe/H] = +0.10 (\textit{right}).}
  \label{fig_cmd_NGC752}
\end{figure}

A color-magnitude diagram of the open cluster NGC752 was obtained with a reddening $E(B - V) = 0.035$ and a distance modulus $(m - M)_{0} = 8.25$ mag (Daniel {\em et al.\/} \cite{daniel94}) for each metallicity (Fig. \ref{fig_cmd_NGC752}). In the left panel of Fig. \ref{fig_cmd_NGC752}, the isochrone of age 1.90 Gyr were constructed with models of metallicity [Fe/H] = -0.15 and masses from 0.80 to 1.59 \msun \ with a step of 0.01 \msun, whereas in the right panel, the isochrone of age 1.49 Gyr was constituted by models of metallicity [Fe/H] = +0.10 and masses from 0.80 to 1.81 \msun \ with a step of 0.01 \msun.

\begin{figure}
  \centering
  \includegraphics[width=5.5cm,height=5.5cm]{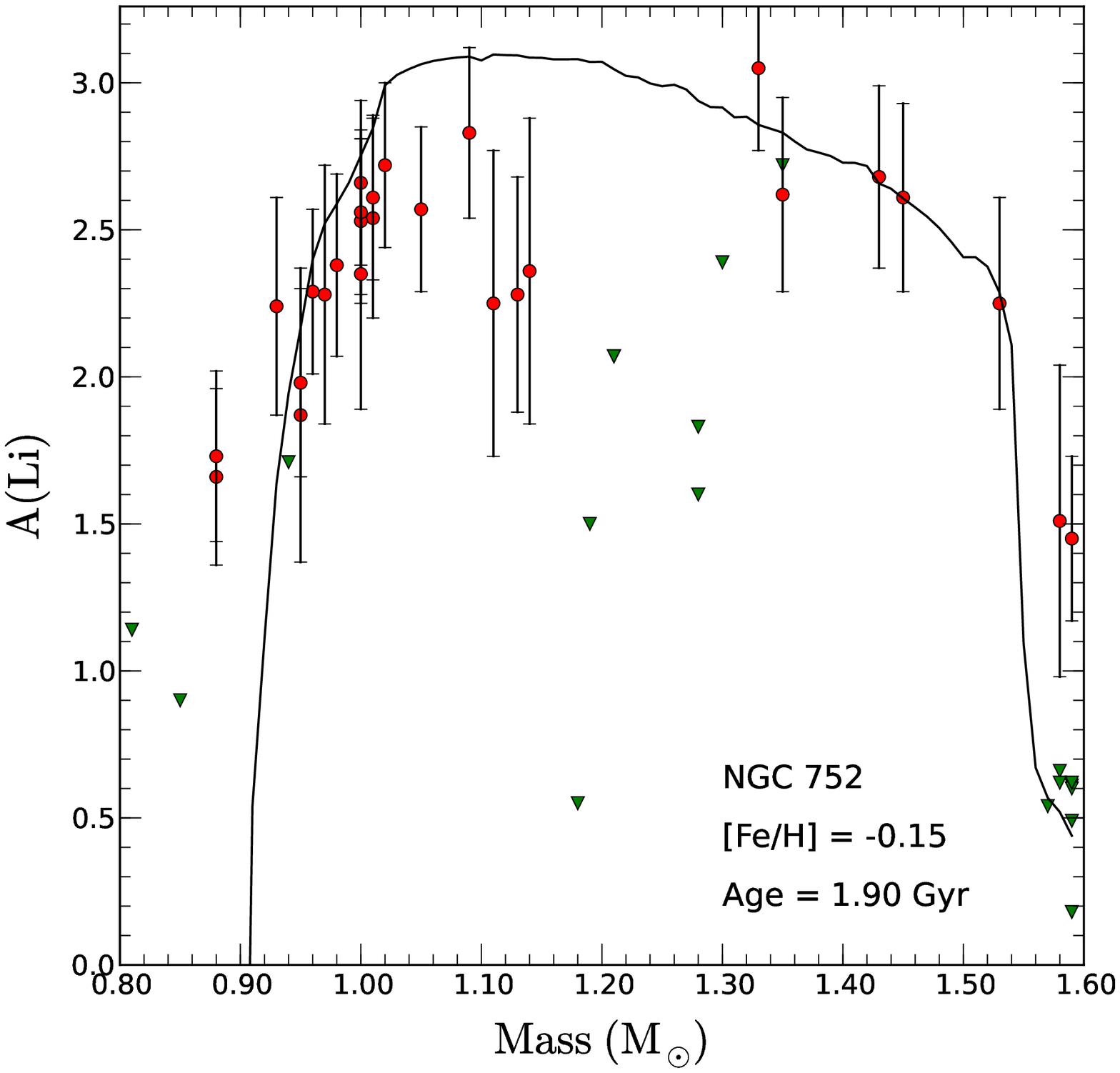}
  \qquad
  \includegraphics[width=5.7cm,height=5.5cm]{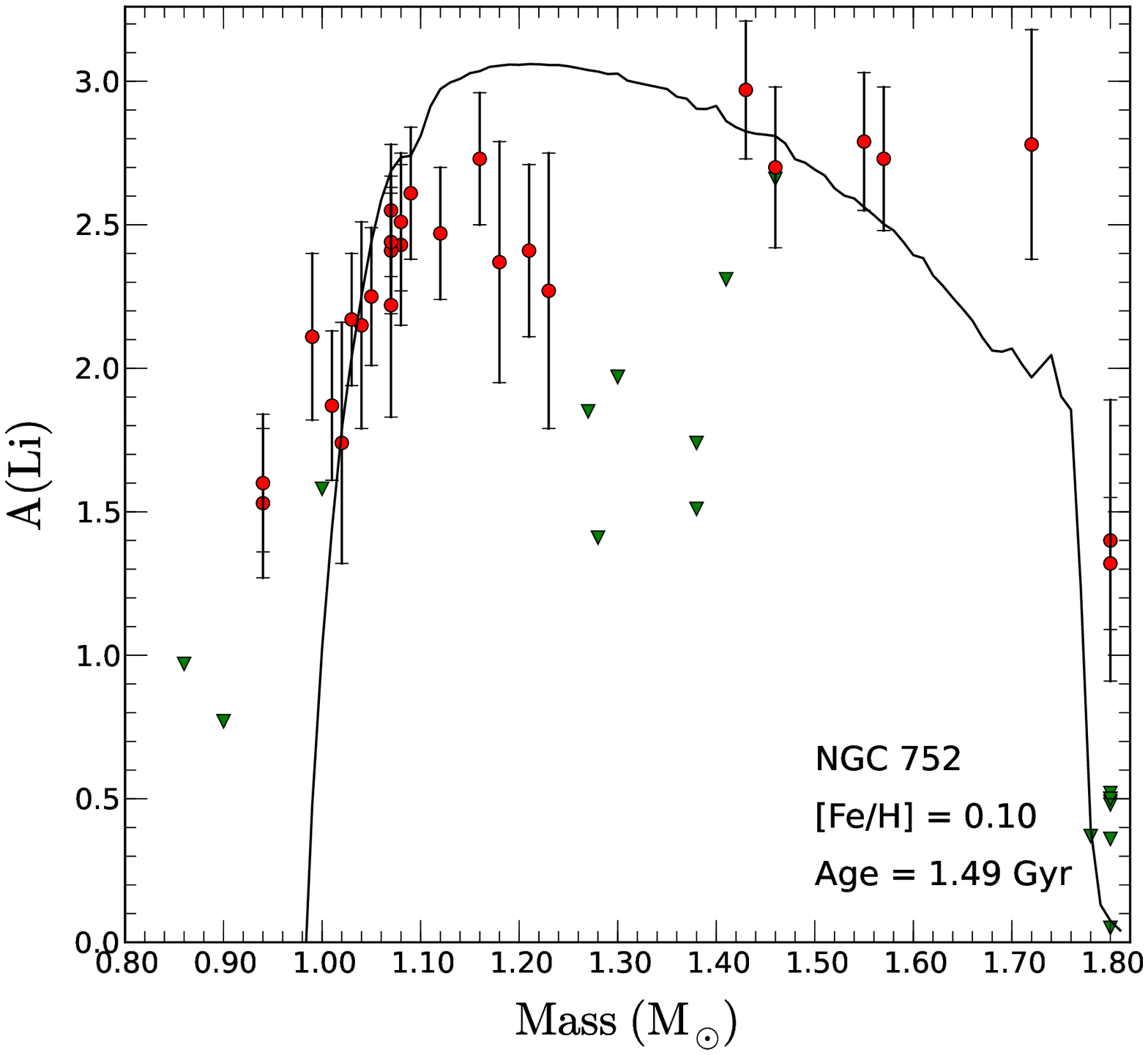}
  \caption{Li abundance as a function of stellar mass for the open cluster NGC752 for metallicities [Fe/H] = -0.15 (\textit{left}) and [Fe/H] = +0.10 (\textit{right}).}
  \label{fig_Li_NGC752}
\end{figure}

In Figure \ref{fig_Li_NGC752}, we show the evolution of Li abundance as a function of stellar mass for each metallicity. In both cases, we have a good agreement for solar-type stars ($M < 1.10$ \msun), and no significant dispersion at this age. However, the Li-dip is not reproduced by our models, which shows a lack of an extra-mixing mechanism, probably the internal gravity waves (Charbonnel \& Talon \cite{charbonnel&talon2005}).

Comparing the right panel of Fig. \ref{fig_M67} and the left panel of Fig. \ref{fig_Li_NGC752}, one notices that the Li-dip in M67  occurs at higher masses as compared to NGC752 if the metallicity given by Daniel {\em et al.\/} (\cite{daniel94}) is adopted for the latter, thus  making Li-dip stars in NGC752 more Li-depleted than M67 members of the same mass. However, if we assume, as reasonable, the metallicity of [Fe/H]=0.10 for NGC752, which was found by Carrera \& Pancino through a high-resolution spectroscopic study, stellar evolution models make stars in this cluster more massive, and the masses of the Li-dip match for the three clusters. This suggests that - while our adopted value for metallicity affects the mass evaluation of any star, therefore also that of Li-dip stars - in reality it is difficult to infer a causal relation between the cluster metallicity and the mass at which the Li-dip occurs.

\section{Li in the open cluster Hyades}
\label{sec_Hyades}

Our sample is composed of 233 stars with photometry from Tycho-2 and parallaxes from Hipparcos (de Bruijne {\em et al.\/} \cite{debruijne2001}). The metallicity [Fe/H] = +0.13 comes from Paulson {\em et al.\/} \cite{paulson2003}. For 67 stars of this sample, we have the Li abundances determined by spectral synthesis by Takeda {\em et al.\/} \cite{takeda2013}.

\begin{figure}
  \centering
  \includegraphics[width=5.5cm, height=5.5cm]{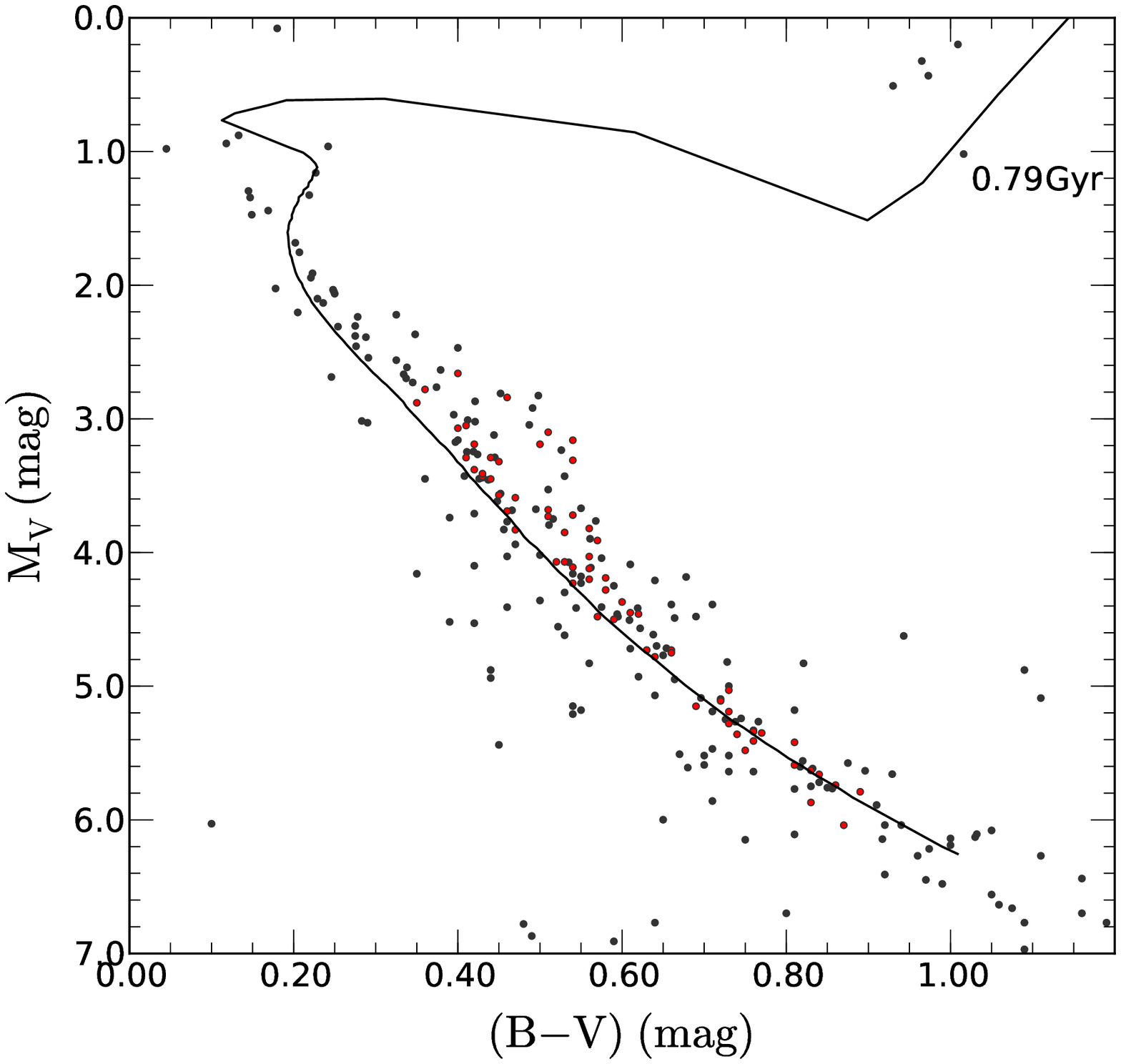}
  \qquad
  \includegraphics[width=5.7cm, height=5.5cm]{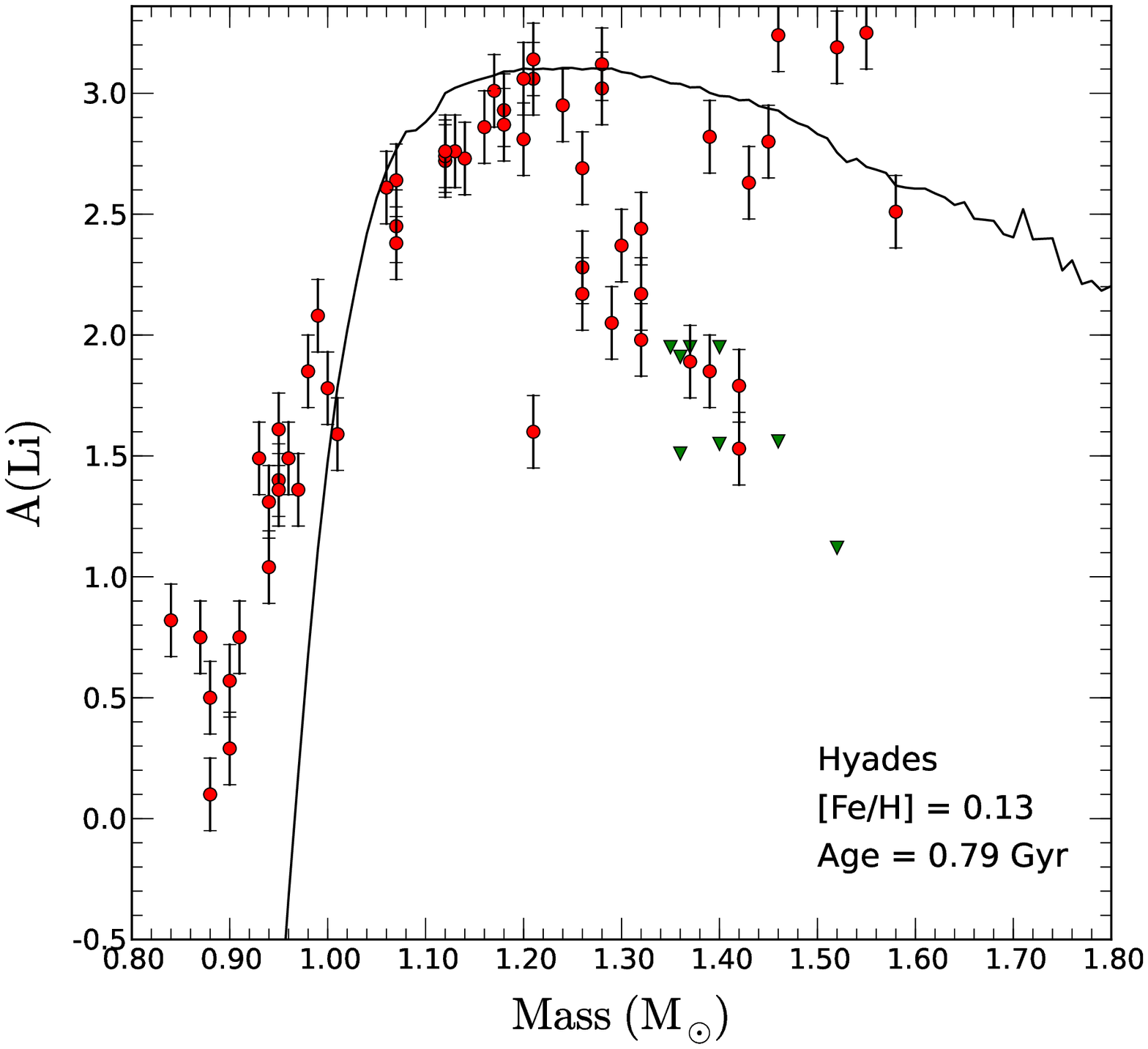}
  \caption{\textit{Left}: Color-magnitude diagram of the open cluster Hyades. The red points represent the 67 stars of our sample for which we have Li abundance determination. \textit{Right}: Li abundance as a function of stellar mass for the Hyades.}
  \label{fig_Hyades}
\end{figure}

The color-magnitude diagram of the open cluster Hyades (left panel of Fig. \ref{fig_Hyades}) is constructed with a reddening $E(B - V) = 0.010$ mag and a distance modulus $(m - M)_0 =$ 3.30 mag. Models of metallicity [Fe/H] = +0.13 and masses from 0.80 to 2.25 \msun \ with a step of 0.01 \msun \ allowed to construct the isochrone of age 0.79 Gyr.

Right panel of Fig. \ref{fig_Hyades} shows the Li abundance evolution as a function of stellar mass for the Hyades. We can see that large destruction of Li occurs very soon in lower mass stars, even if our models destroy too much Li. The masses of stars in the Li-dip of the Hyades match with the masses of stars in the Li-dip of M67 and of NGC752 with the larger metallicity.

\section{Conclusions}
\label{sec_Conclusions}

We have a good qualitative general agreement between observations of Li abundances and models isochrones in the three clusters. However, the Li-dip is not reproduced by the models. We think that the inclusion of internal gravity waves could resolve the problem, even if other extra mixing mechanism cannot be excluded. If we adopt for NGC 752 the metallicity measured through high resolution  spectroscopy ([Fe/H]=0.11, Carrera \& Pancino \cite{carrera&pancino2011}), the position in mass of the Li-dip matches for the three clusters, thus suggesting that NGC752 metallicity is underestimated by photometric measurements.



\begin{thebibliography}{99}
\bibitem[2009]{asplund2009} Asplund, M., Grevesse, N., Sauval, A. J. \& Scott, P. 2009, A\&AR, 47, 481
\bibitem[1995]{balachandran95} Balachandran, S. 1995, ApJ, 446, 203
\bibitem[2010]{baumann2010} Baumann, P., Ram\'irez, I., Mel\'endez, J., Asplund, M. \& Lind, K. 2010, A\&A, 519, A87
\bibitem[2011]{cantomartins2011} Canto Martins, B. L., L\`ebre, A., Palacios, A., \etal \ 2011, A\&A, 527, A94
\bibitem[2011]{carrera&pancino2011} Carrera, R. \& Pancino, E. 2011, A\&A, 535, A30
\bibitem[2011]{castro2011} Castro, M., do Nascimento, J.-D., Jr., Biazzo, K., Mel\'endez, J. \& De Medeiros, J. R. 2011, A\&A, 526, A17
\bibitem[2005]{charbonnel&talon2005} Charbonnel, C. \& Talon, S. 2005, Sci, 309, 2189
\bibitem[1994]{daniel94} Daniel, S. A., Latham, D. W., Mathieu, R. D. \& Twarog, B. A. 1994, PASP, 106, 281
\bibitem[2001]{debruijne2001} de Bruijne, J. H. J., Hoogerwerf, R. \& de Zeeuw, P. T. 2001, A\&A, 367, 111
\bibitem[2009]{donascimento2009} do Nascimento, J.-D., Jr., Castro, M., Mel\'endez, J., \etal \ 2009, A\&A, 501, 687
\bibitem[2013]{donascimento2013} do Nascimento, J.-D., Jr., Takeda, Y., Mel\'endez, J., \etal \ 2013, ApJL, 771, L31
\bibitem[1986]{hobbs&pilachowsky86} Hobbs, L. M. \& Pilachowski, C. 1986, ApJ, 309, L17 
\bibitem[1999]{jones99} Jones, B. F., Fischer, D. \& Soderblom, D. R. 1999, AJ, 117, 330
\bibitem[2000]{montalban&schatzman2000} Montalb\'an, J. \& Schatzman, E. 2000, A\&A, 354, 943
\bibitem[2007]{melendez&ramirez2007} Mel\'endez, J. \& Ram\'irez, I. 2007, ApJ, 669, L89
\bibitem[2010]{melendez2010} Mel\'endez, J., Ram\'irez, I., Casagrande, L., \etal \ 2010, Ap\&SS, 328, 193
\bibitem[2012]{melendez2012} Mel\'endez, J., Bergemann, M., Cohen, J. G., \etal \ 2012, A\&A, 543, A29
\bibitem[2012]{pace2012} Pace, G., Castro, M., Mel\'endez, J., Th\'eado, S. \& do Nascimento, J.-D., Jr. 2012, A\&A, 541, A150
\bibitem[1997]{pasquini97} Pasquini, L., Randich, S. \& Pallavicini, R. 1997 A\&A, 325, 535
\bibitem[2008]{pasquini2008} Pasquini, L.,  Biazzo, K., Bonifacio, P., Randich, S. \& Bedin, L. R.  2008, A\&A, 489, 677 
\bibitem[2011]{pasquini2011} Pasquini, L., Melo, C., Chavero, C., Dravins, D., \etal \ 2011, A\&A, 526, A127
\bibitem[2003]{paulson2003} Paulson, D. B., Sneden, C., Cochran, W. D. 2003, AJ, 125, 3185
\bibitem[1988]{pilachowsky&hobbs88} Pilachowski, C. A. \& Hobbs, L. M. 1988, PASP, 100, 336
\bibitem[1988]{pilachowsky88} Pilachowski, C., Saha, A. \& Hobbs, L. M. 1988, PASP, 100, 474
\bibitem[2002]{randich2002} Randich, S., Primas, F., Pasquini, L. \& Pallavicini, R. 2002, A\&A, 387, 222
\bibitem[2007]{randich2007} Randich, S., Primas, F., Pasquini, L., Sestito, P. \& Pallavicini, R. 2007, A\&A, 469, 163
\bibitem[2004]{richard2004} Richard, O., Th\'eado, S. \& Vauclair, S. 2004, SoPh, 220, 243
\bibitem[2004]{sestito2004} Sestito, P., Randich, S. \& Pallavicini, R. 2004, 426, 809 
\bibitem[2013]{takeda2013} Takeda, Y., Honda, S., Ohnishi, T., \etal \ 2013, PASJ, 65, 53
\bibitem[2009]{xiong&deng2009} Xiong, D. R. \& Deng, L. 2009, MNRAS, 395, 2013
\end{thebibliography}
\end{document}